 	\documentclass[aps,prb,twocolumn,showpacs]{revtex4}

\usepackage{graphicx} 
\usepackage{calc} 
\usepackage{amsfonts}  
\usepackage{amsmath} 
\usepackage{pstricks,pst-grad}
\usepackage{psfrag}
\usepackage{bbold}
\usepackage{bm}

\begin{document}
\newcommand{\kw}{ \bar{k} }
\newcommand{\qq}{  \bar{q}_0  }
\newcommand{\qw}{ \bar{q} }
\newcommand{\G}{ \bar{G} }
\def\sumslashD{\mathop{\sum \kern-1.4em -\kern 0.5em}}
\def\sumslash{\mathop{\sum \kern-1.2em -\kern 0.5em}}
\def\intslash{\mathop{\int \kern-0.9em -\kern 0.5em}}
\def\intslashD{\mathop{\int \kern-1.1em -\kern 0.5em}}
\newcommand{\texteeps}[3][c]{\psfrag{#2}[#1][#1]{\normalsize#3}}
%commandes de Marc
\newcommand{\hc}{{\mathrm{H.c.}}}	%conjugue hermitique
\newcommand{\potloc}{\mathcal U}	%terme de potentiel local
\newcommand{\hamil}{{H}}		%hamiltonien
\newcommand{\paren}[1]{\left(#1\right)}	%pour mettre entre parentheses

\title{Seebeck coefficient in low-dimensional fluctuating charge-density-wave systems}        
 \author{M. Mbodji  and C. Bourbonnais }
\affiliation{  Regroupement Qu\'ebecois sur les Mat\'eriaux de Pointe,
  D\'epartement de physique, Universit\'e de Sherbrooke, Sherbrooke,
  Qu\'ebec, Canada, J1K-2R1 }
  
\date{\today}

\begin{abstract}
We study the role of charge density-wave  fluctuations on the temperature dependence of  Seebeck coefficient in quasi-one dimensional conductors with a Peierls instability. The description of low-dimensional incommensurate charge density-wave fluctuations as obtained by a generalized Ginzburg-Landau approach  for  arrays of weakly coupled chains is embodied  in the numerical solution of the semi-classical Boltzmann transport equation.  The energy and temperature dependence of the scattering time of electrons on fluctuations can then be extracted and its influence on the Seebeck coefficient calculated. The connexion between theory and experiments carried out  on molecular conductors is presented and  critically discussed. 
  \end{abstract} 
\maketitle 

\section{Introduction}

In recent years, the understanding of the role played by   fluctuations  on  transport properties has held particular interest in the study of correlated  low dimensional  electron systems. This is the case  of the considerable attention devoted to the part played by  fluctuations in the origin of linear temperature resistivity which is found, for instance, in the  metallic phase of many unconventional superconductors near their quantum critical point\cite{Coleman05,Cooper09,DoironLeyraud09,Bruin13,Legros19}.  

Although examined  with  less sustained attention, the thermopower or the Seebeck coefficient (SC)  to which
 this work is devoted, is another transport  observable  known to be influenced by fluctuations. When coupled to  electron degrees of freedom,  fluctuations can introduce significant corrections to the free electron gas - linear in temperature $T$ -   prediction of the SC\cite{Paul01,Li07,Kim10,Buhmann13,Buhmann13b,Arsenault13,Mravlje16,Shahbazi16}.
 According to the well known Mott SC formula\cite{Barnard72,Behnia15},  one difficulty with this quantity is that it is influenced by both thermodynamics and effects linked to  the energetic of collisions along the Fermi surface\cite{Barnard72,Behnia04}. Their respective contributions is not easily disentangled and  require a precise knowledge of  fluctuations involved and how they couple  with electrons.
 
  In the framework  of a spin-fermion model\cite{Paul01} and variant of it\cite{Kim10}, nearly critical two-dimensional antiferromagnetic  fluctuations on  thermodynamics  was shown to lead to a logarithmic or power law  enhancement  of specific heat. This was  in turn shown to introduce similar  temperature dependent corrections to the thermodynamic SC component.  A logarithmic  enhancement was found compatible with experiments in some heavy fermions  near a magnetic quantum critical point\cite{Benz99}. 
 As for the influence of the energy and momentum dependence of inelastic scattering time on SC, it  has also been analyzed in different correlated systems. In the case of the single band repulsive Hubbard model in two dimensions, for instance, the combination of linearized Boltzmann theory and the renormalization group method has made it possible to calculate the electron-electron umklapp scattering time, which  contributes distinctively from thermodynamics   to the deviations from a linear-$T$  SC\cite{Buhmann13,Buhmann13b}, as they can be found  in cuprates and pnictides as a function of doping\cite{Laliberte11,Arsenijevic13}.  The same combination of techniques has been used to  calculate the impact of enhanced umklapp scattering by antiferromagnetic fluctuations  on the SC of quasi-one-dimensional organic  metals near a  quantum  critical point that connects  a  spin-density wave state  to superconductivity under pressure \cite{Shahbazi16}. 

In order to further assess the part  played by the collision  dynamics   in the   temperature profile of  SC, it  would be of primary interest to examine  the problem in  well characterized electron systems known to be among the simplest ones dominated by quasi critical fluctuations.  This is the case of  low dimensional metals   undergoing  a Peierls instability. It corresponds to the formation of a lattice distorsion that  is adiabatically connected  with a charge density-wave (CDW) superstructure. In Peierls systems, the coupling between electrons and low energy  CDW fluctuations takes its origin in the electron-phonon interaction, a connection which is well understood.  An additional simplification resides in the fact that  for  many of them the Fermi surface is found to be particularly simple, reducing essentially to  a plain  - one dimensional -  point-like structure.

The organic charge transfer salt  TTF-TCNQ is a well known example  of such Peierls systems  characterized by  a broad temperature domain of CDW fluctuations\cite{Jerome82}. As  a signature of their one-dimensional character, these fluctuations emerge   well above the temperature scale of true long-range order, which is ultimately  triggered by   small but finite  interchain coupling\cite{Pouget76,Khanna77,Kagoshima76}.  The temperature dependence of the SC for this compound in 
various pressure conditions is well documented\cite{Chaikin73,Weyl82}, 
and  will be compared to the theoretical results developed below.

 For this purpose,  we proceed to the numerical integration of the linearized Boltzmann equation   fed by  collisions of electrons on low energy quasi-1D CDW fluctuations. The 1D features  of fluctuations can be calculated accurately by  the functional integral method. The energy profile of the quasi-particles lifetime across the Fermi level can thus be computed and its contribution to the SC obtained as a function of temperature. 
A   connection between theory and experiments in  TTF-TCNQ  can be qualitatively established, indicating that  the influence of CDW fluctuations on the energy dependence of    electron  scattering  can be  a determinant factor in  the  corrections made to the linear temperature dependence of the Seebeck coefficient of  quasi-1D Peierls systems. 
 
 In Sec.~II the properties of low energy  fluctuations  for the  weakly coupled CDW chains problem are reviewed using the functional integral method.  In Sec.~III  the SC  is derived in the framework of the linearized  Boltzmann theory and in the presence of CDW fluctuations.  In Sec.~IV the numerical solution  of the Boltzmann equation is carried out and the calculated Seebeck coefficient is critically compared with the available data  for TTF-TCNQ. We conclude this work in Sec.~V.
  
  %%%%
  \section{Fluctuations of  the Peierls instability}
We consider the following standard minimal Hamiltonian that captures the CDW phase transition for  a set of $N_\perp$ weakly coupled chains of length $L$ 
\begin{align}
\label{ }
H =  & N_\perp \sum_q \hbar \omega_q \big(b^\dagger_qb_q +  {1\over 2}\big) + \sum_{i,k,\sigma}\epsilon_k \, c^\dagger_{i,k,\sigma} c_{i,k,\sigma} \cr 
& {1\over \sqrt{L}} \sum_{i,k,q,\sigma} g\, c^\dagger_{i,k+q,\sigma}c_{i,k,\sigma} (b_q + b^\dagger_{-q})\cr
& + H_\perp
\end{align} 
The first term describes   free  phonons where  $\hbar \omega_q $ is the acoustic phonons spectrum of each chain and $b^\dagger_q$ ($b_q$) is the creation (destruction) operator of phonon of wave vector $q$ along the chains. The second term corresponds to the non interacting electron part of each chain $i$ where  $c^\dagger_{i,k,\sigma}$ ($c_{i,k,\sigma})$ is the creation (destruction) electron operator of wave vector $k$ and spin $\sigma$. We shall consider  the tight-binding electron spectrum 
\begin{equation}
\label{spectrums}
\epsilon_k = 
 -2t_\| \cos kd,  
\end{equation}
which  is  appropriate for the acceptor TCNQ chains of TTF-TCNQ\cite{Berlinsky74}. Here $t_\|$ is the longitudinal hopping and  $d$ is the lattice constant along the chains ($L=Nd$). The third term corresponds to the interaction between electrons and phonons whose electron-phonon matrix element $g$ will be considered   momentum independent for simplicity. In spite of a purely one-dimensional electron structure in  the model, the  possibility of three-dimensional  CDW  order  can be assured by an  interchain part $H_\perp$ for the repulsive   electron-electron (back scattering) interaction between nearest-neighbor chains $i$ and $j$. It  can be written in the form 
\begin{equation}
\label{ }
H_\perp=  g_\perp \sum_{\langle i,j\rangle}\sum_{Q} O_i(Q)^\dagger O_j(Q),
\end{equation}
where $g_\perp>0$. This repulsive term  favors anti-phase CDW ordering in the   directions perpendicular  to the chains; it is expressed in terms of the chain   CDW operator 
$$
 O_i(Q) = 1/\sqrt{L} \sum_{k>0,\sigma} c^\dagger_{i,k-2k_F-Q,\sigma}c_{i,k,\sigma}.
 $$ 
 For the partition function $Z={\rm Tr}\  e^{\beta (H-\mu N)}$, the partial trace over the harmonic phonon degrees of freedom in $H$ generates a retarded  effective  interaction between electrons. Together with  the transverse part $H_\perp$, both terms can be convert  as  a coupling to auxiliary CDW  fields via a Hubbard-Stratonovich transformation\cite{Caron83,Bourbon96}. By carrying out the remaining  trace over electronic degrees of freedom, the  partition function can be put in the following functional integral form
\begin{align}
\label{}
    Z&
      \to \iint [\mathfrak{D}\Delta^*\mathfrak{D}\Delta]  e^{-{\cal H}[\Delta^*,\Delta]},
\end{align} 
 where ${\cal H}[\Delta^*,\Delta]$ is the Landau-Ginzburg Wilson   free energy density functional of the CDW field $\Delta^{(*)}$. Up to quartic order  at  low (Matsubara) frequency  ${\omega_m= 2\pi m k_BT/\hbar}$, small wave vector  deviations  $Q$ from  $2k_F$, and weak  interchain coupling $g_\perp$, the functional can be written in the form 
\begin{align}
\label{functional}
& {\cal H}[\Delta^*,\Delta]   =  \sum_{i,\bar{Q}}  N(\epsilon_F)\big[ \ln T/T_c^0 + \xi_0^2Q^2 + \Gamma_0 |\omega_m| \big] |\Delta_i(\bar{Q})|^2\cr
 & + {k_BT\over L} b\sum_{i,\{\bar{Q}\}} \Delta^*_i(\bar{Q}_1) \Delta^*_i(\bar{Q}_2) \Delta_i(\bar{Q}_3) \Delta_i(\bar{Q}_4)\delta_{\bar{Q}_{1+2}=\bar{Q}_{3+4}}\\
 & + v_\perp \sum_{\langle i,j\rangle} \sum_{\bar{Q}} \Delta^*_i(\bar{Q})\Delta_j(\bar{Q}) \cr
 & \ \ \ \ \ \ \ \ \ \ \ \ \ \equiv {\cal H}_\|[\Delta^*,\Delta] + {\cal H}_\perp[\Delta^*,\Delta].
 \label{functionalb}
\end{align}
The first two terms of (\ref{functional}) define ${\cal H}_\|$, the intrachain part of the functional, whereas the last term describing the interchain Coulomb  interaction between CDW corresponds to the transverse contribution ${\cal H}_\perp$. The functional parameters are given by 
\begin{align}
\label{Tc0}
   T_c^0 &= 1.13 \epsilon_F\, e^{-1/\lambda} \ \ \ \ (\lambda= 2|\bar{g}|^2/\hbar \omega_{2k_F}), \\
   \label{xi0}
   \xi_0 &  = \sqrt{7\zeta(3)\over 16} {\hbar v_F\over \pi k_BT_c^0},\\
   \label{gamma0}
   \Gamma_0 &= { \pi \hbar\over 8k_BT_c^0},\\
    \label{b}
   b & =  {7\zeta(3) \over 16\pi^2} {N(\epsilon_F)\over (k_BT_c^0)^2},\\
   \label{vp}
    v_\perp & = N(\epsilon_F) { \bar{g}_\perp\over\lambda^2},
\end{align}
which in order stand  for the mean field transition temperature $T_c^0$, which fixes the scale of fluctuations of isolated chains, the coherence length $\xi_0$ of $2k_{F}$ electron-hole pairs, the damping constant $\Gamma_0$ of short-range  CDW fluctuations,   the mode-mode coupling constant $b$, and finally the reduced interchain coupling $v_\perp$, expressed as a ratio between interchain and intrachain  couplings. Here $\bar{Q} =(Q,\omega_m)$, $|\bar{g}|^2= |g|^2N(\epsilon_F)$ and $\bar{g}_\perp = g_\perp N(\epsilon_F)$,   $N(\epsilon_F)= (\hbar \pi v_F)^{-1}$ being the density of states at the Fermi level  with the Fermi velocity $v_F={1\over \hbar}\nabla_{k} \epsilon_k|_{k_F}$. 

Since  $v_\perp $ is small,      a perturbative RPA summation  for the quasi-1D dynamic CDW susceptibility after analytic continuation to real frequency $(i\omega_m \to \omega)$, leads  to:
\begin{align}
\label{GRPA}
 G(Q+2k_F,\bm{Q}_\perp + \bm{q}_{\perp0}, \omega)   =   &\,  \langle \Delta^*(Q,\bm{Q}_\perp,\omega)  \Delta(Q,\bm{Q}_\perp,\omega)\rangle  \cr
    = & \,   {G_\|(Q+2k_F,\omega)\over 1 + v_\perp(\bm{Q}_\perp) G_\|(Q+2k_F,\omega)},
\end{align}
where 
\begin{equation}
\label{ }
v_\perp(\bm{Q}_\perp) = -2v_\perp (\cos Q_{\perp y} d_\perp + \cos Q_{\perp z} d_\perp)
\end{equation}
  is the Fourier transform of interchain coupling with the wave vector  $\bm{Q}_\perp=(Q_{\perp y},Q_{\perp z})$ as deviations with respect to the transverse staggered CDW ordering at $\bm{q}_{\perp0}= (\pi/d_\perp,\pi/d_\perp) $; and 
  $$
  G_\|(Q+ 2k_F,\omega) = \langle \Delta^*(Q,\omega)  \Delta(Q,\omega)\rangle_\| 
  $$ 
  is the 1D dynamic CDW susceptibility. We observe that in the static limit, the RPA summation is equivalent to a molecular field approximation of interchain coupling\cite{Scalapino75}.  
  
  To proceed  to the evaluation of $G_\|$, we  first consider  the static part
  \begin{equation}
\label{StaticG}
G_\|(Q+2k_F,\omega=0) = {1\over k_BT} \int \langle \Delta^*(x) \Delta(0)\rangle_\| \ e^{-iQx} dx,
\end{equation}
which   involves the 1D spatial CDW correlation function
\begin{equation}
\label{ }
\langle \Delta^*(x) \Delta(0)\rangle_\| = {1\over Z_\|} \iint [\mathfrak{D}\Delta^*\mathfrak{D}\Delta] \, \Delta^*(x) \Delta(0) \ e^{-\beta {\cal F}_\|[\Delta^*,\Delta]},
\end{equation} 
of a complex CDW order parameter $\Delta(x) = |\Delta(x)|e^{i\varphi(x)}$ with amplitude and phase degrees of freedom. 
In the static limit for ${\cal H}_\| $ in (\ref{functionalb}), this correlator is calculated using the static  Landau-Ginzburg free energy functional
\begin{equation}
\label{ }
{\cal F}_\|[\Delta^*,\Delta] = \int_0^L dx \ \big[ a(T) |\Delta(x)|^2 + c \Big|{d\Delta \over dx}\Big|^2 + b|\Delta(x)|^4\big].
\end{equation}
The parameters of the functional are ${a(T) = a'\ln T/T_c^0}$ ${\simeq a'(T-T_c^0)/T_c^0}$, $a'= N(\epsilon_F)$, $c= N(\epsilon_F) \xi_0^2$, and $b$ is given by (\ref{b}). The calculation can thus be done accurately by the transfer matrix method\cite{Scalapino72,Dieterich76} with the result
\begin{equation}
\label{correlator}
\langle \Delta^*(x) \Delta(0)\rangle_\| =\,  (k_BT_c^0)^2\langle\, |\bar{\Delta} |^2\rangle_\|\  e^{-x/\xi_\|},
\end{equation}
where $\bar{\Delta}= \Delta /k_BT_c^0$  is the normalized CDW order parameter. The CDW static correlations with respect to the wave vector $2k_F$  decay  exponentially as a function of distance with the characteristic  length scale  $\xi_\|$,   the 1D CDW correlation length. The  results of the standard transfer matrix method show that for an incommensurate  two-component order parameter\cite{Dieterich76,Scalapino72}, $\xi_\|$ first grows like $(T-T^*)^{-1/2}$  for temperature ${T< 2T_c^0}$ with an apparent ordering scale $T^*\sim T_c^0/3$, but  ultimately evolves  toward  a $\xi_\|\sim 1/T$  singularity, characteristic of absence of CDW long-range order and dominant phase correlations   at   $T\ll  T^*$.  As for the mean-square of  CDW amplitude  fluctuations $ \langle\,  |\bar{\Delta} |^2\rangle_\| $, it  shows a regular drop as the temperature decreases down to  $T\sim T^*$, and finally rises at lower temperature to reach the saturation mean-field value $ \langle\,  |\bar{\Delta} |^2\rangle_\| \sim 1 $ at $T\ll T^*$.

From (\ref{correlator}), the static 1D susceptibility (\ref{StaticG}) can be put in  the form
\begin{equation}
\label{Gstatic}
G_\|(Q+2k_F,0) = 2k_BT_c^0 {\langle \, |\bar{\Delta} |^2\rangle_\| \over t} {{\xi_0^2\over \xi_\|}\over {\xi_0^2\over \xi_\|^2} + \xi_0^2Q^2 },
\end{equation}
where $t=T/T_c^0$ is the reduced temperature. The denominator  of the above expression shows  the same  $Q$ development   initially obtained  for the quadratic part of the 1D functional ${\cal H}_\|$ in (\ref{functional}). So,  this suggests that a generalization leading to the frequency dependent expression $G_\|(Q+2k_F,\omega)$ can be obtained by simply adding the damping term $-i\Gamma_0\omega$ to the denominator of (\ref{Gstatic}), with the result
\begin{equation}
\label{ }
G_\|(Q+2k_F,\omega) = 2k_BT_c^0 {\langle \, |\bar{\Delta} |^2\rangle_\| \over t} {\xi_\|\over  1 + \xi_\|^2 Q^2  -i \Gamma_\| \omega},
\end{equation}
where the $\Gamma_\| = \Gamma_0 \xi_0^2/\xi_\|^2 $ is the damping constant for CDW correlations of size $\xi_\|$.
  %%%%%%
\section{Linearized Boltzmann approach to Seebeck coefficient}
We approach the problem of scattering of one-dimensional  carriers of charge $e$ on quasi-1D CDW fluctuations in terms of the Boltzmann equation. In presence of the   electric field ${\cal E}$ set up by the thermal gradient $\nabla_x T$ parallel to chains, the stationary form of the transport equation for the Fermi distribution $f(k)$ reads, 
\begin{equation}
\label{Boltzmann}
  \left[ {\partial f_k\over \partial t}\right]_\mathrm{coll} =e {\cal E} \, \nabla_{\hbar k} f_k  - {(\epsilon_k -\mu)\over T} \nabla_x T\, \nabla_{\hbar k} f_k,
\end{equation}    
where $e$ is the electron charge and   $\mu$ is the chemical potential. 

We seek a solution for  $f_k \simeq f^0_k + f^0_k [1-f^0_k] \phi_k$ that is linear in the  deviations $\phi_k$  from  the free equilibrium distribution $f^0_k= (e^{\beta (\epsilon_k-\mu)} +1)^{-1}$. From the Fermi golden rule for the transition rate of electrons on low energy charge CDW fluctuations near $2k_F$, the collision term on the left hand-side of (\ref{Boltzmann}) can be written in the linear form
\begin{align}
\label{PhiColl}
\left[{\partial \phi_k\over \partial t}\right]_{\rm coll} & =  -  {2\over \hbar}  |g|^2\sum_{k'} {1- f^0_{k'}\over 1-f^0_k} (1-\delta_{kk'}) \phi_{k'} \cr
& \times \! \int_{-\infty}^{+\infty} n(\omega) \Im\mathrm{m} \chi (k'-k,\omega) \delta(\epsilon_{k'}-\epsilon_k-\hbar \omega),\cr
\end{align} 
where $n(\omega) =(e^{\beta\hbar \omega} -1)^{-1}$ is the Bose factor and $\Im\mathrm{m} \chi (k'-k,\omega)$ is the imaginary part of the retarded phonon Green function. The latter can be connected to the CDW retarded response by replacing the phonon operators  by their CDW  macroscopic configurations, namely $b^{(\dagger)}\to \Delta^{(*)}/(2g) $. Thus from the fluctuation-dissipation theorem, the imaginary part of intrachain CDW response can be related to the strength of  fluctuations obtained from (\ref{GRPA}). At low frequency, one finds   
\begin{align}
\label{ }
\Im\mathrm{m} & \chi(Q+2k_F,\omega) =    {\beta \hbar \omega \over 2|g|^2}\cr
\times
 & {1\over N_\perp^2}\sum_{\bm{Q}_\perp} \Re\mathrm{e} \,G(Q+2k_F,\bm{Q}_\perp+\bm{q}_\perp^0,\omega),
 \end{align}
which is strongly peaked at $\hbar \omega \sim \epsilon_{k'}-\epsilon_{k}$. Performing  the frequency integration in (\ref{PhiColl}), the linearized Boltzmann equation can be written in the following integral form 
\begin{align}
\label{LinBoltzmann}
 \left[{\partial \phi_k\over \partial t}\right]_{\rm coll} =  &\, k_B \beta^2 v_k (\epsilon_k -\mu) \nabla_xT -e \beta {\cal E} v_k \cr
  = & -{\cal L}\phi_k  = -\sum_{k'} {\cal L}_{kk'} \phi_{k'},
\end{align}
from  which we define the collision operator  
\begin{widetext}
\begin{align}
\label{L}
 {\cal L}_{kk'} &  =   \ {2 \over t^{2}}{1 \over LN_\perp^2}\sum_{k',\bm{Q}_\perp} {1-f_{k'}^0\over 1-f_k^0}\hbar^{-1}\left({\epsilon_{k'}-\epsilon_k}\right) n[(\epsilon_{k'}-\epsilon_k)/\hbar] {\langle \, |\bar{\Delta} |^2\rangle_\| } { \xi_\|}\cr
 & \times\left(  {   1 + \xi_\|^2(k'-k)^2 - Y_\perp(\bm{Q}_\perp)\over\big[ 1 + \xi_\|^2(k'-k)^2 - Y_\perp(\bm{Q}_\perp)\big]^2 + (\Gamma_\|/\hbar^2) (\epsilon_{k'}-\epsilon_k)^2}  \right) (1-\delta_{kk'}).   
\end{align}
\end{widetext}
The transverse part is given by 
\begin{equation}
\label{Y}
Y_\perp(\bm{Q}_\perp) ={4 \alpha\over t}  {\langle\, |\bar{\Delta}|^2\rangle_\| }  { \bar{g}_\perp\xi_\|\over\lambda^2\xi_0} \big(\cos(Q_{\perp y}d_\perp)+ \cos(Q_{\perp y}d_\perp)\big)
\end{equation}
and $\alpha \simeq 2.9$.

It is convenient to write the solution for $\phi_k$ as the sum of two terms 
\begin{align}
\label{}
  \phi_k  = & - {\cal L}^{-1} k_B \beta^2 v_k (\epsilon_k -\mu) \nabla_xT  + {\cal L}^{-1}e \beta {\cal E} v_k  \cr
    \equiv &  -\phi^T_k + \phi_k^{\cal E}
\end{align} 
where $ {\cal L}^{-1}$ is the inverse of the collision operator. The parameters entering in the collision operator are those of the spectrum in (\ref{spectrums}); together with the scale $T_c^0$, these fix the Ginzburg-Landau parameters (\ref{Tc0}-\ref{b}) for 1D fluctuations leading to  $\xi_\| $ and $ \langle\, 
|\bar{\Delta}|^2\rangle_\|$. The small interchain coupling (\ref{vp}) in (\ref{Y}) is fixed in order to get the desired true $T_c\ll T_c^0$.
%%%%%
\subsection{Seebeck coefficient}
To obtain the  expression for the Seebeck coefficient, we start from the relation  of the 1D electric current density along the chains, which to linear order in $\phi$ reads
\begin{align}
\label{}
   j_\| = &\, {2 e\over L} \sum_k v_k f_k     \simeq\, {2 e\over L} \sum_k v_k f^0_k(1-f^0_k)(  \phi_k^{\cal E}-\phi^T_k). 
\end{align}
Using the normalized quantities 
$$
\bar{\phi}_k^T= {\phi^T_k\over k_B\beta^2 v_k (\epsilon_k-\mu)\nabla_x T}
$$
and 
$$
\bar{ \phi}_k^{\cal E} =  {\phi_k^{\cal E}\over e\beta v_k {\cal E}}.
$$
The longitudinal current density can then be written in the form
\begin{equation}
\label{ }
j_\| = K_{11} {\cal E} -K_{12} \nabla_x T.
\end{equation}
For open circuit conditions, $j_\|=0$, and the  expression for the SC  follows,
\begin{align}
\label{Qa}
Q= & { {\cal E}\over \nabla_xT} = {K_{12}\over K_{11}}\cr
 = &\, {2e L^{-1} \sum_k k_B\beta^2 v_k^2 (\epsilon_k-\mu) f_k^0(1-f_k^0) \bar{\phi}_k \over 2e^2 L^{-1}\sum_k \beta v_k^2 f^0_k(1-f_k^0)  \bar{\phi}_k  },
\end{align}
where $\bar{\phi}^{{\cal E},T} (\equiv \bar{\phi}_k) $ obeys the single equation
\begin{equation}
\label{phibar}
{\cal L} \bar{\phi}_k = \sum_{k'} {\cal L}_{kk'} \bar{\phi}_{k'} =1,
\end{equation}
whose solutions  can be considered as the scattering time of carriers as a function of $k$ or the energy $\epsilon_k$. Since at low temperature, the product of Fermi distribution factors in (\ref{Qa}) is strongly peaked at the Fermi level $\mu$, a Sommerfeld development leads to the Boltzmann type of expression 
\begin{align}
\label{Qf}
Q = & \, {\pi^2\over 3} {k_B^2 T\over e} \left[ {d\ln N(\epsilon_k)\over d\epsilon_k}\Big|_\mu + 2{d\ln v_{k}\over d\epsilon_k}\Big|_\mu + {d\ln \bar{\phi}_{k}\over d\epsilon_k}\Big|_\mu\right]\cr
= & \, Q_0 + Q_c.
\end{align} 
Here $Q_0$ stands as the band or thermodynamic contribution to the SC, which corresponds to the sum of the first two terms of (\ref{Qf}). From the form of spectrum  in (\ref{spectrums}), one recovers  the known result  
\begin{equation}
\label{Q0}
Q_0=  
- \displaystyle{{\pi^2 \over 3}{k_B \over|e|} {\epsilon_F\over 4t_\|^2- \epsilon_F^2}} k_BT
\end{equation}
for the tight-binding electrons in one dimension, where $\epsilon_F= |\epsilon_{k_F}|$.

The last term   $Q_c$ is linked to the dynamics of collisions  at the Fermi level; it is computed from the  numerical solution of (\ref{phibar}) using the expression (\ref{L}) for the collision operator\cite{Note2SeebeckCDW}. This will lead to the energy or momentum profile of the collision time across the Fermi level. 

\section{Numerical results}
\subsection{One dimension}
%%%%%%%%%%%%%%%%%%%%%
    \begin{figure}
    %\begin{psfrags}
   % \psfrag{?}{$\displaystyle{?}$}
   \includegraphics[width=8.0cm]{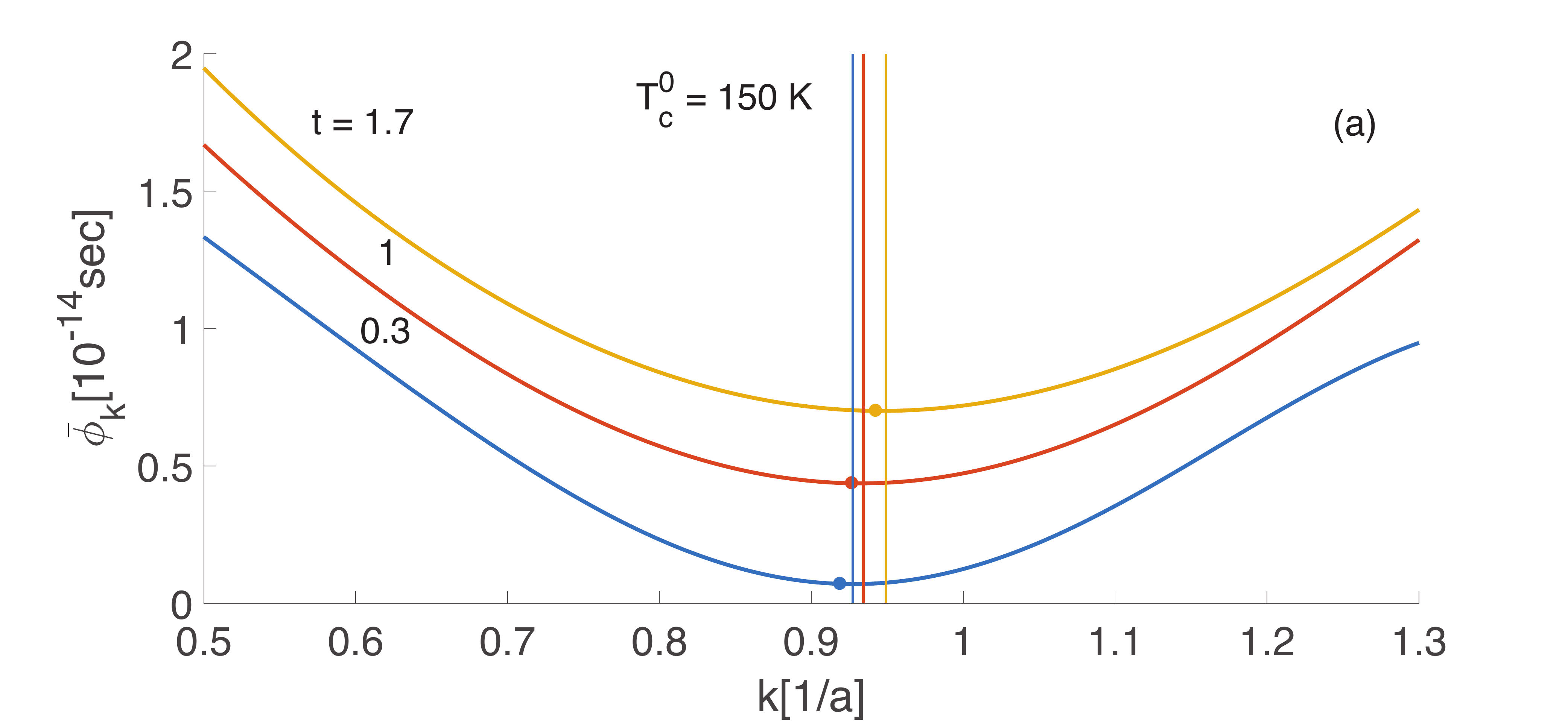}\vskip 0.5cm
     \includegraphics[width=8.0cm]{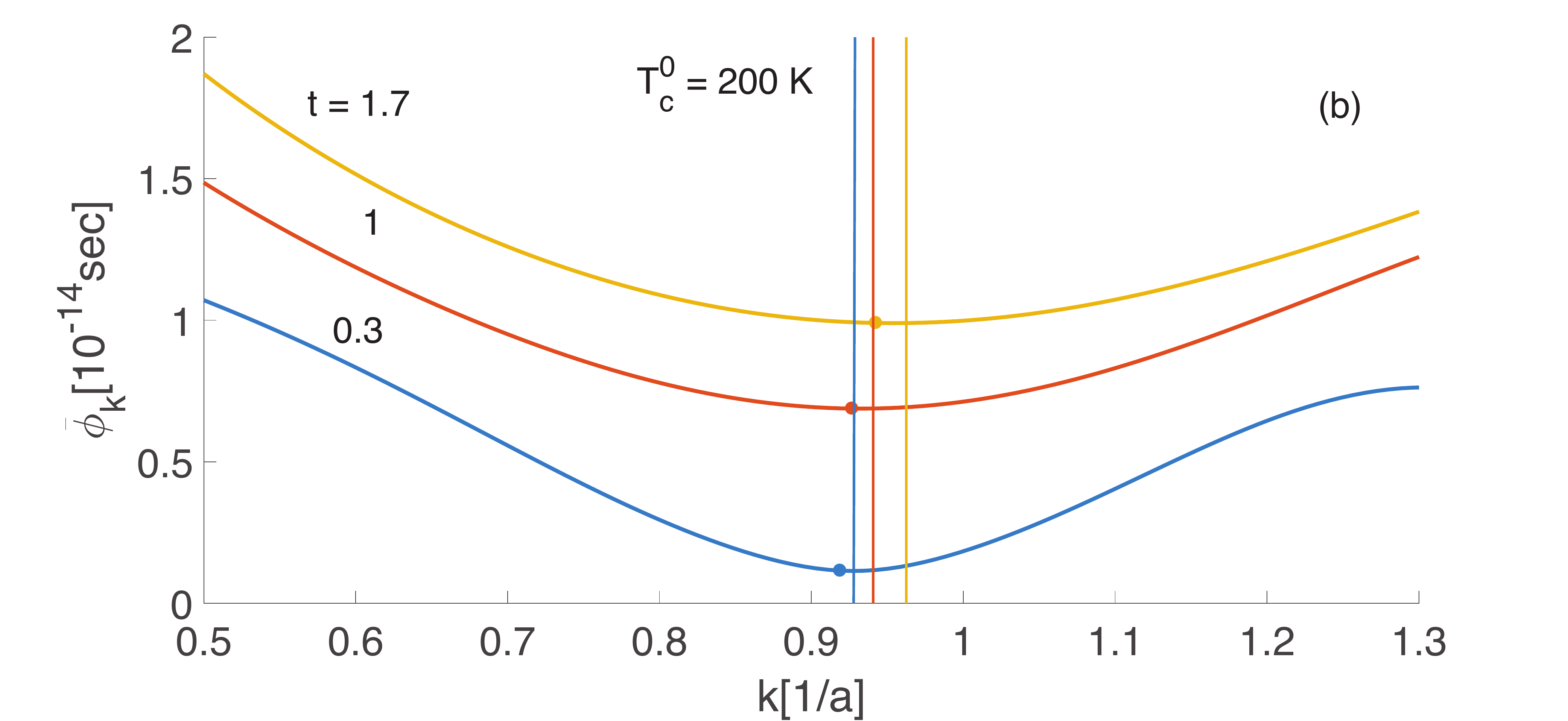}
  \caption{Collision time as a function of the wave vector $k$ in the neighborhood of the Fermi point $k_F(t)$ at different reduced  temperatures $t= T/T_c^0$ in (a) the one-dimensional case for $T_c^0=150$~K and (b) for coupled chains at $T_c^0=200$~K. The location of $k_F(t)$ as a function of temperature  is indicated by vertical lines and the minimum by points. }
  \label{Phi} 
  %\end{psfrags}
   \end{figure}
%%%%%%%%%%%%%%%%%%%%%%%
It is instructive to first consider the results of the above equations for the Seebeck coefficient in the purely 1D limit where $Y_\perp=0$, namely when the interchain Coulomb term $g_\perp $ is put to zero.  The parameters used  in the calculations are those typically found for the TCNQ chains of TTF-TCNQ in normal pressure conditions\cite{Jerome82,Khanna77,Berlinsky74}, namely $\rho=0.59$ for the incommensurate electron concentration taken at low temperature, $t_\|=0.11 {\rm eV}$ for the hopping along the chains and a set of  $T_c^0$  for the scale of 1D CDW fluctuations. The   temperature interval for all calculations  is fixed  to $t\le 2$, namely where  $\xi_\| \ge \xi_0$ and short-range CDW order can be considered as relevant. 

In Figure~\ref{Phi} we show the  $k>0$  dependence of the collision time $\bar{\phi}_k$ near the Fermi point $k_F(t)$  for different reduced temperatures $t$.  The small variation of $k_F(t)$ with temperature is taken into account from the corresponding shift of the chemical potential $\mu(t)$, which is linked by the relation, $\rho= 2/N \sum_k f_k^0$, between the electron concentration $\rho$ and the fermi distribution factor.  
From the Figure, we observe at all $t$ the existence of a minimum in the collision time close to $k_F(t)$, indicating enhanced scattering by CDW fluctuations  between $k$ and $k-2k_F$ states. The minimum deepens as $t$ decreases and fluctuations effects increase.  However, the true minimum of $\bar{\phi}_k$ is slightly shifted downward from $k_F(t)$, which reflects the electron-hole asymmetry normally expected for electron carriers for which the collision time increases with energy. This asymmetry introduces a finite positive slope of $\partial \bar{\phi}_k/\partial k\big|_{k_F(t)}$ ($\equiv \hbar v_{k} \partial \bar{\phi}_k/\partial \epsilon_{k} \big|_{\mu(t)}$), which  according to (\ref{Qf}) will yield corrections to the linear-$T$ prediction  $Q_0$  for  the SC.  

\begin{figure}
    %\begin{psfrags}
   % \psfrag{?}{$\displaystyle{?}$}
   \includegraphics[width=8.5cm]{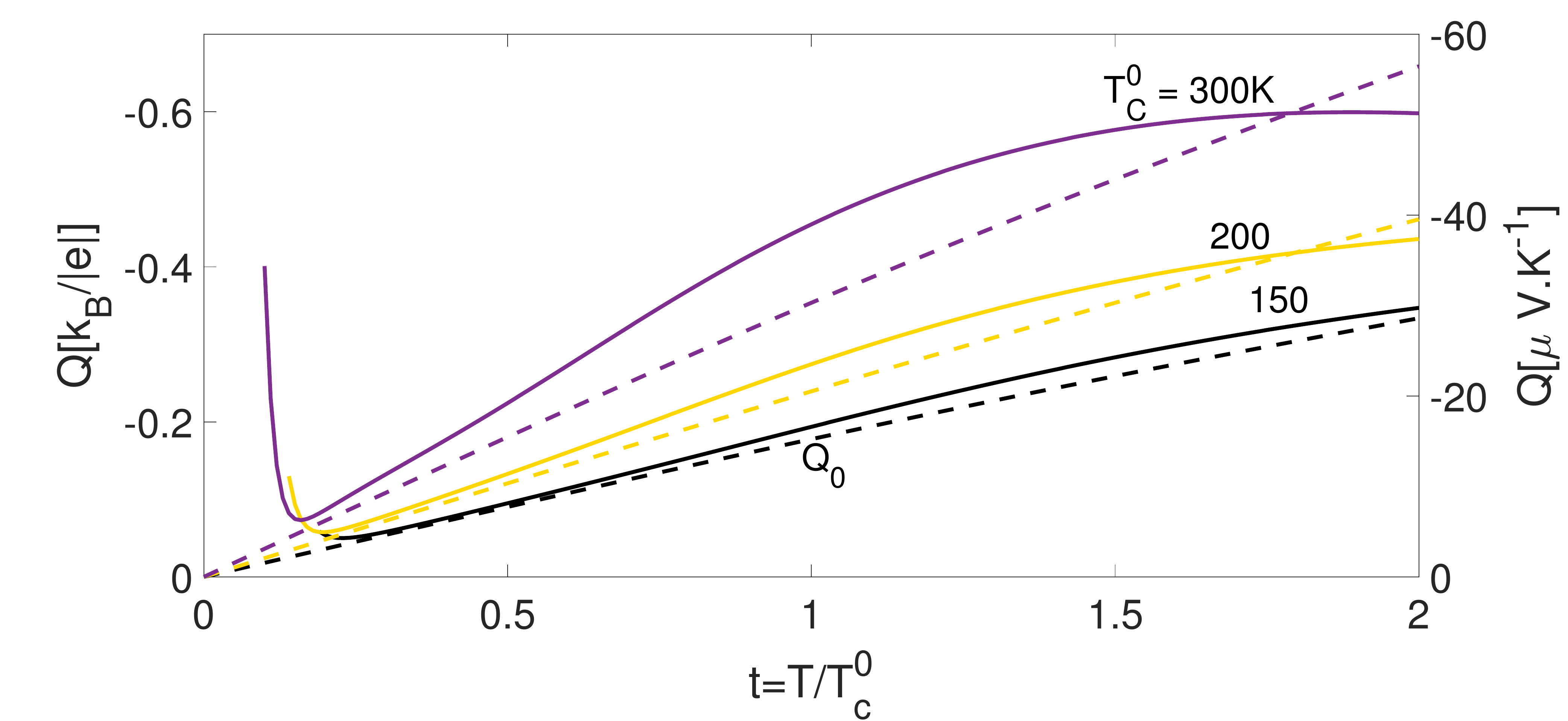} 
  \caption{Calculated one-dimensional  Seebeck coefficient as a function of the reduced temperature $t=T/T_c^0$ for different fluctuation scales $T_c^0$. The dashed lines correspond to  the free electron, tight-binding,  band contribution $Q_0$. }
  \label{Seebeck1D} 
  %\end{psfrags}
   \end{figure}

The resulting temperature profile of the 1D SC (\ref{Qf}) is shown in Figure \ref{Seebeck1D}  for different scales of $T_c^0$.  We observe that  in all cases  a sublinear  temperature dependence  in the high temperature domain where amplitude fluctuations are important; it becomes more pronounced with the size of  $T_c^0$. A change of regime is found below $t\sim1$, where phase fluctuations of the CDW order parameter emerge; it evolves toward an effective but enhanced   $T$-linear behavior that follows the empirical form\cite{Behnia04}
\begin{equation}
\label{zeta}
Q \approx  Q_0(1+ \zeta)
\end{equation}
with the constant $\zeta>0$ that increases the slope of $Q$ with $T_c^0$.  Following  Refs \cite{Barnard72,Behnia04}, this  is  compatible with  a power law dependence $\bar{\phi}_k \sim |\epsilon_k|^{-\zeta} $  of the collision time upon the tight-binding energy, which increases as $ \epsilon_k$ grows (or $v_k$ increases). At very low temperature  where $t\ll 1$,   $\xi_\|$ and phase correlations become very large which deepen the minimum in $\bar{\phi}_k$, resulting in an upturn of the   SC. 
%%%%%%%%%%%%%%%%%%%%%
%%%%%%%%%%%%%%%%%%%%%%%

\subsection{Quasi one-dimensional case} 
\label{Q1D}
We now turn to the situation where the interchain coupling $g_\perp$ is finite and long-range CDW ordering  emerges at $T_c \ll T_c^0$. According to the RPA expression (\ref{GRPA}), when evaluated at $\bm{q}_0=(2k_F,\pi/d_\perp,\pi/d_\perp)$, $g_\perp$ can be  fixed in order  to give a singularity at  $t_c=T_c /T_c^0\simeq 0.36\  (0.27)$ for $T_c^0=150$~K (200~K) and $t_\|=0.11$~eV,  which corresponds to the $T_c\simeq 54$~K typically found for TCNQ chains, using the known low temperature incommensurate electron concentration $\rho=0.59$ in normal pressure conditions \cite{Jerome82}. With these sets of parameters, the variation of the collision time as a function of $k$ across the Fermi point  $k_F(t)$ is shown in Figure~\ref{Phi}~(b) at different temperatures above $t_c$  for ${T_c^0= 200}$~K. From the Figure, we observe that for temperatures not too close to  $t_c$, the $\bar{\phi}_k$   and positive slopes at $k_F(t)$ essentially coincide  with those found in  the 1D case for the same $T_c^0$, indicating that in this temperature domain the influence of interchain coupling is weak and collisions are primarily governed by 1D fluctuations.   As for the dependence on the temperature  scale for fluctuations, there is an overall downward shift of $\bar{\phi}_k$  with $T_c^0$ and  an increase in slope at $k_F(t)$.  
%%%%%%%%%%%%%%%%%%%%%
    \begin{figure}
    %\begin{psfrags}
   % \psfrag{?}{$\displaystyle{?}$}
   \includegraphics[width=9.0cm]{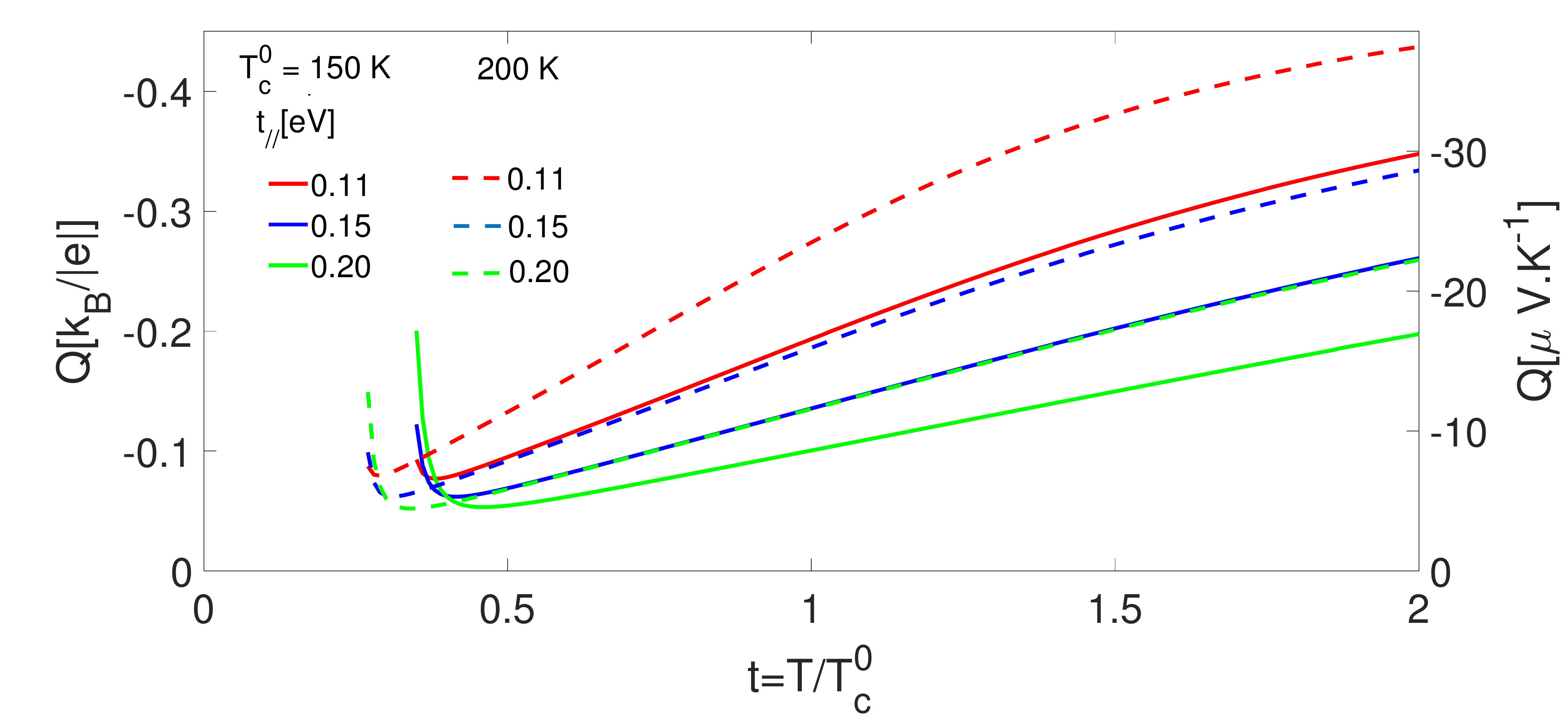} 
  \caption{Calculated three-dimensional  Seebeck coefficient as a function of the reduced temperature $t=T/T_c^0$ for different fluctuation scales $T_c^0$ and band parameters $t_\|$.  }
  \label{Seebeck3D} 
  %\end{psfrags}
   \end{figure}

The  temperature dependence  of the SC is shown in Figure~\ref{Seebeck3D} for different $T_c^0$ and  band parameters $t_\|$. At $t_\| =0.11$~eV, we verify that for both values of $T_c^0$, the $Q$ temperature dependencies for coupled chains   display a sublinear variation that increases with $T_c^0$ above $t\sim 1$ and an enhanced linear-$T$ behavior below. Both features  essentially  coincide  with those found in  Figure~\ref{Seebeck1D}  for the 1D case, and this over a large part of the temperature interval. It is only at the approach of  $t_c$ where CDW  correlations  are singular  that  an upturn in the absolute value of  SC is found. 

Also shown in the Figure is the influence of an increase of the band parameter $t_\|$. We observe that as $t_\|$ grows, all the above features for $Q$ soften. According to (\ref{Q0}), an increase in the electron bandwidth will  decrease the slope of  $Q_0$, as normally expected; it will also reduce  the collision term $Q_c$ in  (\ref{Qf})  through  an increase of the Fermi velocity or equivalently a decrease in the density of states at the Fermi level.

We close the section by examining the combined  influence of varying both $t_\|$ and the electron concentration  $\rho$ following the relation $\delta \ln \rho \simeq  0.36\, \delta \ln t_\|$, which correlates the average variations of  electron concentration $(\delta \ln \rho/\delta P\simeq 0.9\%/{\rm kbar})$ and hopping $(\delta \ln t_\|/\delta P\simeq 2.5\%/{\rm kbar})$ found in a  compressible system like TTF-TCNQ, namely when pressure is applied and  temperature is lowered \cite{Jerome82}.  
%%%%%%%%%%%%%%%%%%%%%
    \begin{figure}
    %\begin{psfrags}
   % \psfrag{?}{$\displaystyle{?}$}
  \includegraphics[width=9.0cm]{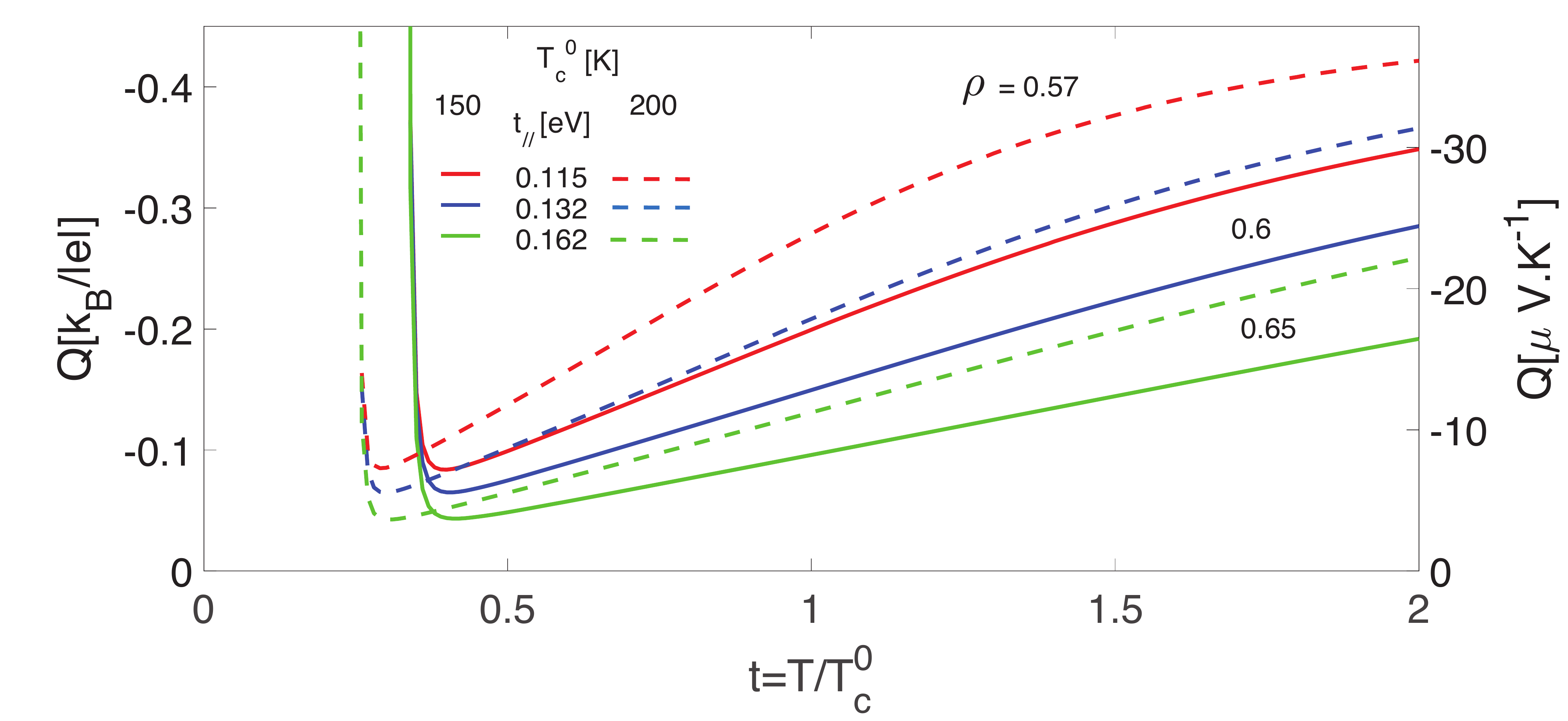} 
  \caption{Calculated three-dimensional  Seebeck coefficient as a function of the reduced temperature $t=T/T_c^0$ for different electron concentrations  $\rho$, band parameters $t_\|$,  and scales $T_c^0$ for fluctuations.   }
  \label{SeebeckRho} 
  %\end{psfrags}
   \end{figure}
%%%%%%%%%%%%%%%%%%%%%%%
This is displayed in Figure~\ref{SeebeckRho} for different $T_c^0$ compatible with x-ray data\cite{Pouget76,Khanna77,Kagoshima76}. Thus an increase of electron concentration $\rho$   in the electron carrier sector  $\rho<1$ for the spectrum (\ref{spectrums}), produces a decrease of the SC, which adds to that of $t_\|$ shown previously in Figure~\ref{Seebeck3D}.  As $\rho$ grows from its initial average value of 0.57 at ambient pressure, the Fermi point $k_F$ is shifted upward and with it the size of the Fermi velocity. The resulting decrease  of the density of states at the Fermi level  weakens both $Q_0$ and   $Q_c$ contributions. Thus as  $\rho$ and $t_\|$ grow, the sublinear  $T$ dependence of  the Seebeck coefficient  is then gradually suppressed; it is replaced by a linear temperature variation  with a  reduced slope congruent with the empirical expression (\ref{zeta}), but with a small $\zeta$. This indicates  that at sufficiently large $\rho$ and $t_\|$ corrections coming from the collision term becomes relatively small. This holds  outside the critical domain of the transition  where fluctuations become large and produce an upturn in $|Q|$  which is more pronounced with increasing  $\rho$ (See also note \cite{Note1SeebeckCDW}). 

\section{Connection with experiments in TTF-TCNQ}

%%%%%%%%%%%%%%%%%%%%%
    \begin{figure}
    %\begin{psfrags}
   % \psfrag{?}{$\displaystyle{?}$}
   \includegraphics[width=9.0cm]{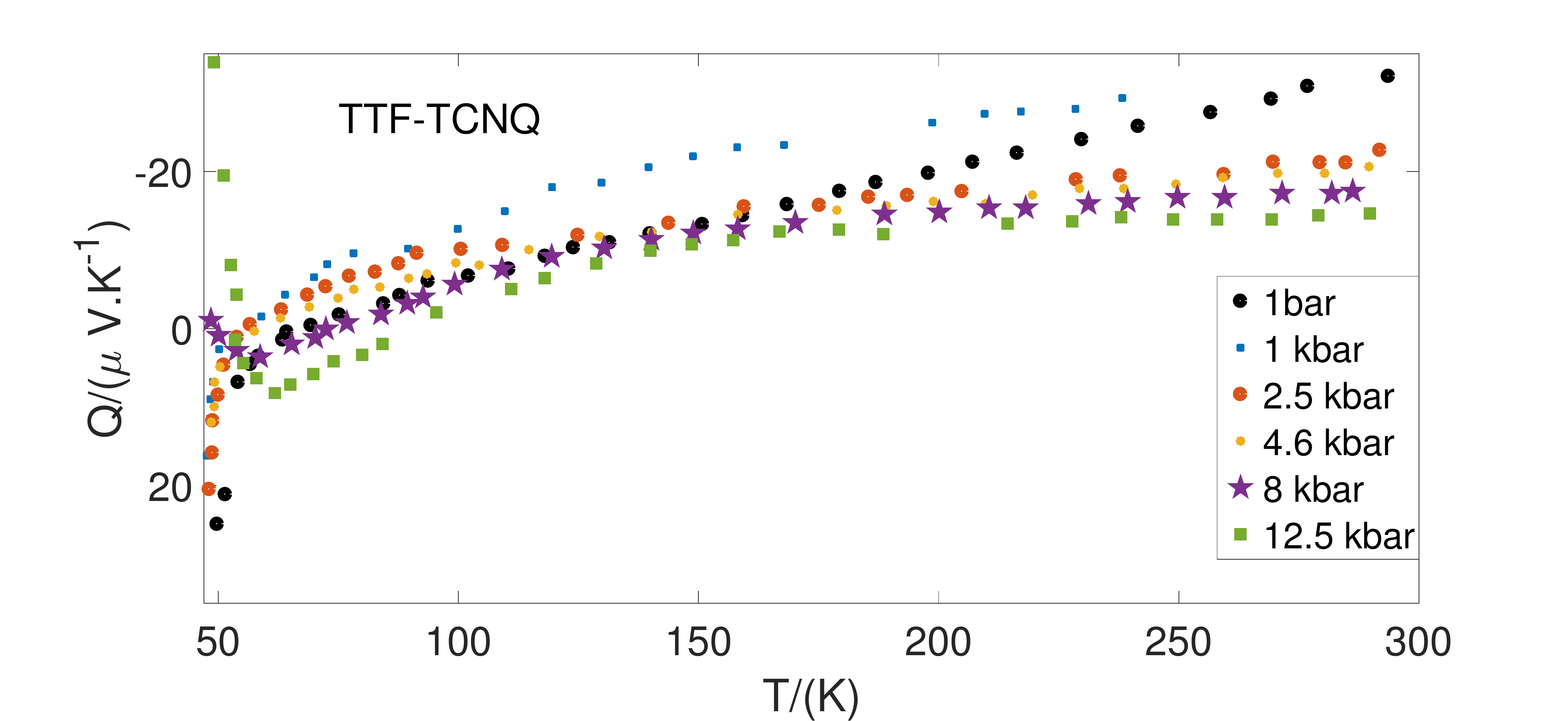} 
  \caption{Seebeck coefficient data as a function of temperature  in TTF-TCNQ  at ambient and different applied pressures. The data are reproduced down to the close proximity of $T_c$ for the TCNQ chains. After Refs \cite{Chaikin73,Weyl82}.  }
  \label{SeebeckExp} 
  %\end{psfrags}
   \end{figure}
%%%%%%%%%%%%%%%%%%%%%%%
We now compare the above theoretical results to existing experimental data for the SC in TTF-TCNQ. Although this two-chain charge transfer salt has  both electron  and hole like type of carriers pertaining to TCNQ and TTF chains respectively, the conductivity of TCNQ chains ($\sigma_{\rm TCNQ}$) is known to largely dominate  that of TTF  ($\sigma_{\rm TTF}$)\cite{Jerome82,Weyl82}. Following the expression of the Seebeck coefficient for two type of carriers,
$$
Q = {Q_{\rm TCNQ}\sigma_{\rm TCNQ} +   Q_{\rm TTF}\sigma_{\rm TTF}\over \sigma_{\rm TCNQ} + \sigma_{\rm TTF}},
$$
as weighted by the respective conductivities of the chains.  
Since $\sigma_{\rm TCNQ}\gg \sigma_{\rm TTF}$ holds\cite{Weyl82}, it will be assumed to be so for the SC, namely  that $Q \approx  Q_{\rm TCNQ}$  in the whole temperature range $ T_c< T< 2T_c^0 $ considered in the present calculations.

We reproduce in Figure~\ref{SeebeckExp}, the SC temperature dependence in the chain direction  for TTF-TCNQ, as  obtained by Chaikin {\it et al.}\cite{Chaikin73} at ambient pressure (black dots, $P= 1$bar) and for temperature down to $T_c$. We first see that the range of $-30\mu\cdot$V/K  reached by the observed $Q$ at ambient temperature is congruent with the one found  at $t\sim 2$ in the calculations of Figures~\ref{Seebeck3D} and \ref{SeebeckRho}  using the $t_\|=0.11$~eV and   the interval of electron concentration  $\rho=0.57-0.59$ and   the x-ray scale $T_c^0= 150-200$K  for the TCNQ chains at ambient pressure  \cite{Pouget76,Khanna77,Kagoshima76}. Considering  the size of the band contribution $Q_0$ in Fig.~\ref{Seebeck1D}, this is compatible with  relatively small but  finite corrections coming from  the collision term.  The data show a change of behavior  for  $T\sim 150$~K below which  a  more rapid decrease of $Q$ is observed. The progressive increase in the slope of $Q$   as temperature is lowered is  however more pronounced than found in the calculations of  Figures~\ref{Seebeck1D}-\ref{SeebeckRho}   at  $t<1$. Experiments rather reveal that $Q$ evolves toward a change of sign as $T_c$  is approached from above, whereas calculations show an enhanced linear-$T$ behavior before an  upturn in $|Q|$  at the approach of  $t_c$. The difference may take its origin in a shift of the position of the Fermi level as temperature is lowered and fluctuations become large; this would change the sign of $\partial \bar{\phi}_k/\partial k\big|_{k_F(t)}$ at the Fermi level and makes the collision dynamics hole-like instead of electron-like (See note \cite{Note1SeebeckCDW}). One may also be tempted to attribute  the emergence of a positive $Q$ to  the contribution of hole carriers  coming from the TTF chains, as fluctuations and the onset of a pseudogap  on  TCNQ chains grow. However,  other transport measurements on TTF-TCNQ have shown  that at ambient pressure the Hall coefficient   becomes more negative  in the same temperature range, pointing to   the still dominant contribution of electron carriers of TCNQ chains\cite{Cooper77,Jerome82}.

We can now look in Figure~\ref{SeebeckExp} at the temperature dependence of Seebeck coefficient for TTF-TCNQ  under  pressure, as  obtained by Weyl {\it et al.}\cite{Weyl82}  up to 12.5~kbar.     From the Figure we see that pressure steadily decreases the amplitude and the slope of the Seebeck coefficient in the high temperature region, where it appears only weakly temperature dependent at high pressure values. Following the example of the situation found at ambient pressure, a  more rapid decrease of the amplitude of the Seebeck coefficient  is observed at all pressures below ${T\sim 150}$~K, which evolves toward a change of sign of $Q$ at  the approach of the critical temperature $T_c$. This incursion into the positive region  becomes less pronounced with pressure and 
gives way to  a sharp upturn  back to high negative values close to $T_c$.

For the comparison with the present calculations, we  first note that in  this pressure range, the    critical temperature $T_c$ observed for   TCNQ chains varies  little under pressure and remains relatively close to the ambient pressure value of 54~K \cite{Friend78,Weyl82}; $T_c$ has  then been kept approximately constant in the calculations. Regarding the fluctuations scale   $T_c^0$, there is no  available data concerning its  pressure evolution. However, given the empirical relation $T_c\sim T_c^0/3$, which is found in practice when both temperature scales are accessible, $T_c^0$  will   then be taken to fall in the same range $T_c^0\sim 150-200$~K used  at ambient pressure and in   Figures~\ref{Seebeck3D} and \ref{SeebeckRho}. 
The comparison  must also go through a readjustment of the electron concentration $\rho$ and the longitudinal hopping $t_\|$ with pressure. Both quantities increase following the modification of the charge transfer and intermolecular distances under pressure and lowering temperature due to the high  compressibility of a system like TTôF-TCNQ\cite{Megtert79,Berlinsky74,Kagoshima76,Jerome82}. As discussed in  Sec.~\ref{Q1D}, in our constant-volume calculations, we will consider their average variations under pressure   and temperature which are congruent with  the relation   $\delta \ln \rho \simeq  0.36\, \delta \ln t_\|$ established for TTF-TCNQ \cite{Weyl82,Megtert79,Cooper79}. This is the variation used in  Fig.~\ref{SeebeckRho} and  which  cover   about 13 kbar of pressure.

Under pressure the observed  decline in the amplitude and slope of $Q$    are  relatively well caught by the calculations of Figure~\ref{SeebeckRho}  in the high temperature domain, although at the highest pressures  the experimental temperature dependence shown in Fig.~\ref{SeebeckExp} becomes rather weak, making hard the distinction, within experimental accuracy, between a nearly constant behavior and    $T$-linearity with a strongly reduced slope. As for the change of regime in  the Seebeck coefficient observed for all  pressures below  150~K or so, it   is  only apparent in the calculations of Fig.~\ref{SeebeckRho} up to intermediate values of $\rho$ and $t_\|$, above which the contribution of collisions is predicted to be small, excepted sufficiently close to $T_c$. 
The change of sign of $Q$ observed above  $T_c$  under pressure is then not reproduced by calculations, at least  along the line  $\rho(t_\|)$ of the $(\rho, t_\|)$  plane used in the calculations of Figure~\ref{SeebeckRho} (See note \cite{Note1SeebeckCDW}).  In the present framework, this would indicate that the collision term $Q_c$ remains sufficiently important in practice when approaching the  critical temperature region; as previously emphasized for ambient pressure results, the change of    sign could result  from reversing the positions of the Fermi level $k_F(t)$ and the minimum of the scattering time in Fig.~\ref{Phi}.   Finally, as pressure increases, the sharpening of the upturn of  $|Q|$  due to critical scattering close to  $T_c$  is fairly well reproduced by calculations of Fig.~\ref{SeebeckRho}.

  %%%%%%%%%%%%%%%%%%%%%%%%%%%%%%%%%%%%%
 
 \section{Concluding remarks}
 In this work    the temperature dependence of the Seebeck coefficient has been calculated for  correlated quasi-one-dimensional metals dominated by charge-density-wave fluctuations.  A functional integral method was used for the description of  low energy CDW fluctuations ascribed to the precursors of the  Peierls superstructure. As the source of inelastic scattering for carriers, these fluctuations  were embodied  in  the numerical solution of the linearized Boltzmann equation  which  governs the dynamics of scattering   time near the Fermi surface.      The related corrections to the linear-$T$ dependence  of the  SC could thus be obtained and assessed as a function of the strength  of fluctuations.  The analysis was carried out for  typical parameters of low dimensional organic metals,  like  the two-chain compound TTF-TCNQ, for which the negatively charge carriers of TCNQ chains  undergo  a Peierls instability against the formation of a  CDW superstructure.

The  calculations show that the size of SC corrections linked to the scattering dynamics  are congruent with those seen in experiments for TTF-TCNQ,    reproducing  certain experimental features in the temperature dependent SC at ambient and finite pressures.  However, calculations  fail  to reproduce  the incursion of SC towards positive values at the approach of the critical CDW temperature, at least for   the model parameters  that  most realistically fit to TTF-TCNQ \cite{Note1SeebeckCDW}.  

As to  the origin of the change of sign, one can invoke the possibility that CDW   fluctuations also  affect     the thermodynamic term $Q_0$  of the SC  in (\ref{Qf}). Such  corrections   were ignored in the present work.  Fluctuations can modify the energy dependence of the  density of states and carrier velocity  near the Fermi  level and then transform  an electron-like band into a hole-like one. However,  although the presence of an electron pseudo-gap induced by fluctuations  is visible on the   TCNQ chains, as shown by the corresponding Knight shift  spin susceptibility\cite{Takahashi84b}, it is likely to be too small an effect to lead the needed modifications in the energy dependence of  electron velocity that would   yield an  effective change in the  sign  of  the  carriers.   This is borne out by  measurements of the Hall coefficient,   which still displays    strongly negative values in the temperature domain  where   SC becomes positive\cite{Cooper77}. This brings us back to  the energy dependent scattering time considered throughout  this work as the most plausible  cause of this change of sign.  Remember that for compressible molecular  compounds such as TTF-TCNQ, both electron bandwidth and band-filling evolve with decreasing temperature\cite{Jerome82}. These   variations that cannot be  taken into account accurately in  constant-volume calculations like those developed above. Such  variations may be  responsible for the small shift of the   Fermi point  needed to transform electron type scattering into hole one.
  
The applicability of the   present  theory of the Seebeck coefficient to other quasi-1D fluctuating Peierls systems  is rather straightforward,  requiring only   modifications of electronic band and fluctuations scales.    Among them, let's mention for instance the two-chain Peierls compounds TMTSF-DMTCNQ\cite{Andrieux79a} and TTF[Ni(dmit)$_2$]$_2$\cite{Kaddour13,Kaddour14},  for which the  temperature and pressure dependence of the Seebeck coefficient  in the fluctuating metallic phase presents similar features to those  discussed in the present work for    TTF-TCNQ.

 \acknowledgments
 The authors thank Patrick Fournier and Andr\'e-Marie Tremblay for numerous discussions and comments on several aspects of this work. 
 C.B thanks the National Science and Engineering Research Council  of Canada (NSERC) and the R\'eseau Qu\'eb\'ecois des Mat\'eriaux de Pointe (RQMP)  for financial support.

  \bibliography{/Users/cbourbon/Dossiers/articles/Bibliographie/articlesII.bib}

\begin{thebibliography}{41}
\bibitem[{Not({\natexlab{a}})}]{Note2SeebeckCDW}
\bibinfo{note}{{The numerical solution of the linear collision operator ${\cal
  L} \bar{\phi}_k=1$ is carried out by dropping momentum conserving processes
  from the equation, which do not contribute to the Seebeck coefficient}}.
\bibitem[{Not({\natexlab{b}})}]{Note1SeebeckCDW}
\bibinfo{note}{{It is worth noting that the critical fall off of $Q$ towards
  large negative values is not systematically found for all $t_\|$ and $\rho<1$
  when $T\to T_c$. There are actually ranges of values in the $(\rho<1,t_\|)$
  plane that depart slightly from the line, $\delta \ln \rho \simeq 0.36\,
  \delta \ln t_\|$, used in Figure~\ref{SeebeckRho} for TTF-TCNQ, and for which
  $Q$ changes sign and becomes hole-like with large positive values as $T\to
  T_c$}}.
\end{thebibliography}
\end{document}